# Tunable Localized Charge Transfer Excitons in a Mixed Dimensional van der Waals Heterostructure

Mahfujur Rahaman[1], Emanuele Marino[2,3], Alan G. Joly[4], Seunguk Song[1], Zhiqiao Jiang[2,5], Brian T. O'Callahan[4], Daniel J. Rosen[5], Kiyoung Jo[1], Gwangwoo Kim,[1] Patrick Z. El-Khoury[4], Christopher B. Murray[2,5], and Deep Jariwala[1]

[1]Department of Electrical and Systems Engineering, University of Pennsylvania, PA 19104, USA

[2]Department of Chemistry, University of Pennsylvania, PA 19104, USA

[3]Dipartimento di Fisica e Chimica, Università degli Studi di Palermo, Via Archirafi 36, 90123 Palermo, Italy

[4] Physical and Chemical Sciences Division, Pacific Northwest National Laboratory, Richland, WA 99352, USA

[5] Department of Materials Science and Engineering, University of Pennsylvania, PA 19104, USA

**Abstract**

Observation of interlayer, charge-transfer (CT) excitons in van der Waals heterostructures (vdWHs) based on 2D-2D systems has been well investigated. While conceptually interesting, these charge transfer excitons are highly delocalized and spatially localizing them requires twisting layers at very specific angles. This issue of localizing the CT excitons can be overcome via making mixed dimensional vdWHs (MDHs) where one of the components is a spatially quantum confined medium. Here, we demonstrate the formation of CT excitons in a 2D/quasi-2D system comprising $MoSe_2$ and $WSe_2$ monolayers and CdSe/CdS based core/shell nanoplates (NPLs). Spectral signatures of CT excitons in our MDHs were resolved locally at the 2D/single-NPL heterointerface using tip-enhanced photoluminescence (TEPL) at room temperature. By varying both the 2D material, the shell thickness of the NPLs, and applying out-of-plane electric field, the exciton resonance energy was tuned by up to 120 meV. Our finding is a significant step towards the realization of highly tunable MDH-based next generation photonic devices.

**Introduction**

Interlayer excitons (ILXs) are composed of Coulomb bound electron and hole (*e-h*) pairs confined in two different spatially separated quantum wells that are coupled together electronically. Owing to large spatial separation of *e-h* pairs, ILXs possess much longer lifetimes (1 – 3 order of magnitude higher) than the direct excitons of individual QWs[1,2]. This allows ILXs to be subsequently explored for strongly correlated condensed matter phenomena such as Bose-Einstein condensates as well as in excitonic, and photonic devices[3,4]. Experimental observation of ILXs was first reported in coupled GaAs/AlGaAs QWs and later in various III-V and II-VI QW heterostructures[5]. However, the very small exciton binding energy (few meV) of conventional 3D semiconductor QW heterostructures limited the progress of this field to cryogenic measurements[6].

The recent emergence of both structural as well as electronic variety in 2D materials has opened new opportunities to study ILXs. Van der Waals heterostructures (vdWHs) composed of several combinations of distinct 2D materials, especially transition metal dichalcogenides (TMDCs), allow the formation of ILXs with remarkably high binding energies (100 – 350 meV)[7]. Hence, it is possible to observe ILXs in such vdWHs at room temperature (RT), which has made it an intense research topic in recent years[8,9]. ILXs formed in 2D/2D systems are generally delocalized in the 2D plane and require a specific twisting angle between the participating monolayers to create localized excitons in the 2D landscape[10,11]. As a result, forming localized ILXs in a twisted heterobilayer (HB) can function as quantum dot-like (QD) confined potentials which unlock exciting opportunities towards high-performance semiconducting lasers, single photon emitters, entangled photon sources, and tunable exotic quantum phases of matters[4,12,13]. Despite the recent great efforts of spatially confining ILX in HBs with precisely controlled angles, imperfection in crystals, challenges with sophisticated sample preparation, and the repulsive interaction between the confined excitons keep the localization process far from ideal both in terms of spectral lines and spatial extent[14,15].

In this context, mixed dimensional heterostructures (MDH) composed of 2D materials on one side and 0D or spatially confined materials on the other side can be an attractive option for the creation of localized ILXs. Due to the van der Waals nature of the interface formed between 2D and 0D or spatially confined materials, MDHs favor similar charge transport phenomena analogous to all-2D vdWHs, when formed with type-II band alignment[16–19]. Therefore, it is predicted that MDHs can also emit ILX-like excitons, which are known as hybrid or charge transfer (CT) excitons[20,21]. Additionally, reduced dimensionality of one of the materials can introduce arbitrary spatial and energy confinement as well as additional degrees of freedom at the interfaces to tune electronic properties of MDHs[20]. Hence, in contrast to delocalized ILXs in all-2D vdWHs, CT excitons formed in MDH heterointerfaces should be localized along the reduced dimensional materials in the out-of-plane direction. This leads to the possibility of investigating and manipulating localized CT excitons in the 2D landscape of the respective MDH. Moreover, owing to the differences in the density of states and dielectric screening environments on either side of the heterostructure, the mechanism of the CT exciton formation and the consequent parameters that can control it may be fundamentally different compared to all-2D systems[22]. Therefore, MDHs present a new platform to investigate charge-transfer physics and subsequent exciton formation in MDHs, and will have a broader technological impact on many device applications[23,24].

In this work, we report on the observation of CT excitons in MDHs composed of 2D transition metal dichalcogenides (TMDCs) and colloidal semiconducting $CdSe/Cd_xSZn_{1-x}S$ core/shell nanoplates (NPLs). Even though these nanoplates are colloidal semiconducting nanocrystals, their density of states more resemble a step-like quasicontinuum similar to a 2D electronic system[25]. Therefore, these nanoplates are known as quasi-2D (Q2D) systems. We adopt tip- enhanced photoluminescence (TEPL) nano-spectroscopy to resolve the spectral signature of CT excitons from a single NPL/2D heterointerface. Taking advantage of large tunability of the band structure as a function of shell thickness of CdSe/Cd$_x$SZn$_{1-x}$S based core/shell NPLs and combining them with monolayer MoSe$_2$ and WSe$_2$ we are able to tune the CT exciton up to 120 meV. Our work presents primary experimental evidence of the presence of CT excitons with large tunability in a MDH system.

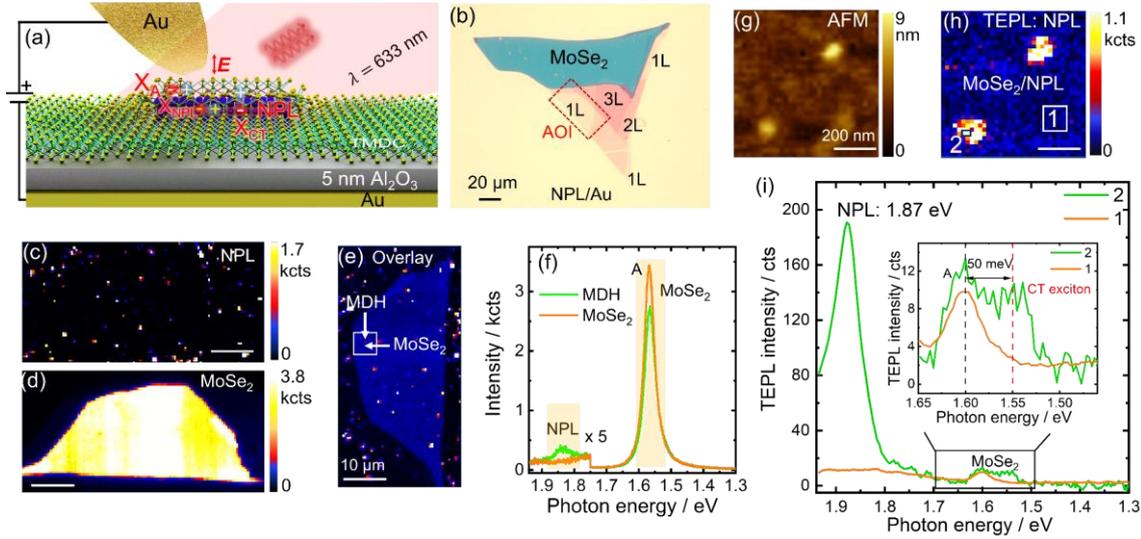

**Figure 1. Micro- and nano-optical characterization of CT excitons in MDHs.** (a) Schematic representation of TEPL measurements of the MDHs containing TMDC monolayers (MoSe$_2$ and WSe$_2$) on top of CdSe/CdS$_x$Zn$_{1-x}$S core/shell NPLs on an Au (or Al$_2$O$_3$/Au) substrate. For the electric field dependent study an out-of-plane bias was applied through the tip and metal substrate. (b) Optical image of one representative MDH device investigated in this work. Area of interest (AOI) is outlined by a dashed rectangle. (c), (d) Far-field PL intensity map of NPL and MoSe$_2$ acquired for the AOI region. (e) PL intensity overlay image using (c) and (d) showing the MDH interface formation on a MoSe$_2$ monolayer. Scale bar is 10 μm. (f) two representative far-field PL spectra of MoSe$_2$ and MDH acquired from two nearby pixels as marked in the overlay image. NPL Spectral regions are multiplied by 5 for better visibility. Orange shades are the spectral regions for which PL maps were created for NPL and MoSe$_2$ respectively. (g), (h) AFM topography and corresponding TEPL intensity map of NPLs on monolayer MoSe$_2$ acquired simultaneously. (i) Two representative TEPL spectra averaged over the rectangle areas marked by 1 and 2 in the TEPL image.

**Results and Discussions**

Fig. 1a presents a schematic of the TEPL configuration used to characterize MDHs in this study. The MDHs containing monolayer $MoSe_2$ (or monolayer $WSe_2$) and $CdSe/Cd_xSZn_{1-x}S$ core/shell NPLs have three different excitons: two in-plane excitons from the TMDCs and NPL respectively and one out-of-plane CT exciton across the MDH interface as schematically presented. A gold tip was used to excite the plasmonic field underneath using 633 nm excitation. Fig. 1b shows an optical image of one of the representative MDH samples studied in this work. Details of the MDH device fabrication, NPLs synthesis and characterization can be found in the method section and the supplementary information section I. Far-field PL intensity maps created for NPLs and $MoSe_2$ and their overlay image for the area of interest (AOI) region marked in the optical image (Fig. 1b) are shown in Fig. 1c-e respectively. The representative far-field-PL spectra for both MDH and monolayer $MoSe_2$ are displayed in Fig. 1f. The orange shades are the spectral region for which the NPL and $MoSe_2$ PL maps were created in Fig. 1c, d. As can be seen in Fig. 1e, the MDHs form at multiple locations between NPLs and $MoSe_2$. Wherever they form an electronic contact, MDHs emit CT excitons as revealed by TEPL. However, it is challenging to resolve CT excitons in the far-field-PL configuration due to the close proximity of this peak to the A exciton of $MoSe_2$ and the large probing cross-section of the far-field PL geometry (~ 0.2 $\mu m^2$) compared to a very small CT exciton emitting area (limited by the spatial extent of NPLs: 6 x $10^{-4}$ $\mu m^2$). These factors ultimately, lead to very weak CT exciton signals in the far-field PL spectroscopy geometry, (see supplementary information section II for more details).

The situation can be changed by introducing TEPL, which excites/emits signal locally under the tip apex with high spatial resolution. Fig. 1g,h show atomic force microscope (AFM) and corresponding TEPL intensity images of NPLs on $MoSe_2$ respectively. Our sub-20 nm spatial resolution was enough to resolve CT exciton from a single NPL/2D MDH interface (see supplementary information Fig. SI-2ii). Despite the excellent sensitivity of TEPL (both enhancement and spatial resolution), the large extent of the 2D plane can still introduce an additional challenge in resolving spectral features from the MDH areas since the TEPL signal needs to overcome a large far-field background (see supplementary information section III). Hence, all the TEPL measurements were treated with far-field background subtraction to resolve the CT excitons in MDHs clearly. Two representative TEPL spectra, one averaged from the NPL region (black rectangle area) and the other from MoSe$_2$ (white rectangle area) marked in Fig. 1h can be seen in Fig. 1i. It is Important to note that MDH spectrum is averaged over 4 pixels (pixel size is 20x20 nm) corresponding to two NPLs (see table S1 in the supplementary section I) making it noisier than the MoSe$_2$ spectra (averaged over 42 pixels). We can clearly observe three PL features in the MDH spectrum in Fig. 1i. Among them, two main excitonic features, NPL PL and MoSe$_2$ A exciton peak are observed at 1.87 eV and 1.60 eV respectively. The third PL peak is observed 50 meV below the MoSe$_2$ A exciton peak in the MDH spectra. We assigned this peak as the CT exciton via *e-h* recombination from NPL to MoSe$_2$ and will provide further evidence to our claim in the following sections.

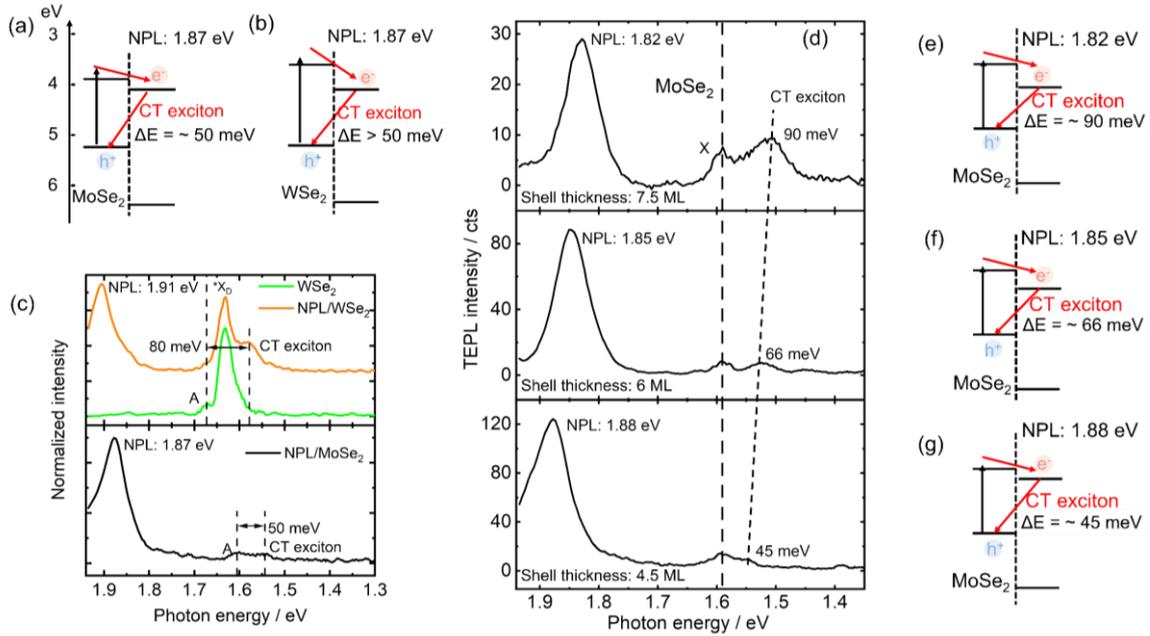

**Figure 2. Band alignment engineered tunability of CT excitons in MDH.** (a), (b) Band alignment diagrams at 2D/Q2D interfaces studied in this work. Values are taken from the refs[26,27]. (c) TEPL spectra acquired for the WSe$_2$/NPL system. For comparison TEPL spectra of bare WSe$_2$ and MoSe$_2$/NPL of Fig. 1i are also plotted. (d) Tuning of CT excitons via NPL shell thickness variation for NPL/MoSe$_2$ system. (e) – (g) Schematic illustration of band alignment for CT exciton tuning via NPL shell thicknesses.

To understand the CT mechanism and the consequent exciton formation at the MDH interface under investigation, we analyse the band diagram using the band values available in the literature for both the semiconductors in the MDH[26,27]. The band alignment presented in Fig. 2a predicts photoexcited electron transfer from MoSe$_2$ to NPLs and then *e-h* recombination from NPL CB to MoSe$_2$ VB, which can be observed as CT exciton in our system. We followed two approaches to confirm the origin of this emission peak. The first one is via changing the 2D material as shown in the band alignment diagram of Fig. 2b. Since the band edges of monolayer WSe$_2$ move higher in energy compared to monolayer MoSe$_2$, it should create CT excitons of smaller energy (larger energy offset) when paired with NPLs of the same bandgap. TEPL spectra acquired for NPL/WSe$_2$ system presented in Fig. 2c demonstrates this hypothesis. For comparison, TEPL spectra of bare WSe$_2$ acquired from a nearby area and the NPL/MoSe$_2$ spectrum of Fig. 2d is also plotted together. The main excitonic feature (bright exciton, A) of WSe$_2$ is observed at 1.66 eV as a shoulder to the strong dark exciton, X$_D$ around 1.62 eV in our TEPL spectra. Even though dark excitons in WSe$_2$ are not permitted in far-field geometry, they can be observed in TEPL at RT due to the strong coupling between the out-of-plane exciton dipole moment and the plasmonic field in the nano-cavity[28,29]. Nevertheless, most importantly, the CT exciton peak can be observed at 80 meV below the WSe$_2$ A exciton. Hence, from Fig. 2c, it is clear that the NPL/WSe$_2$ interface creates CT excitons with larger band offset than the NPL/MoSe$_2$ interface. Note that the slight deviation of the bandgap of NPL is due to thickness variation (a consequence of NPL synthesis process) of the NPL.

As a second approach, we also test the possibility of tuning the CT exciton energy via changing the NPL shell thickness. As predicted in the literature, quasi-type II CdSe/CdS based core/shell

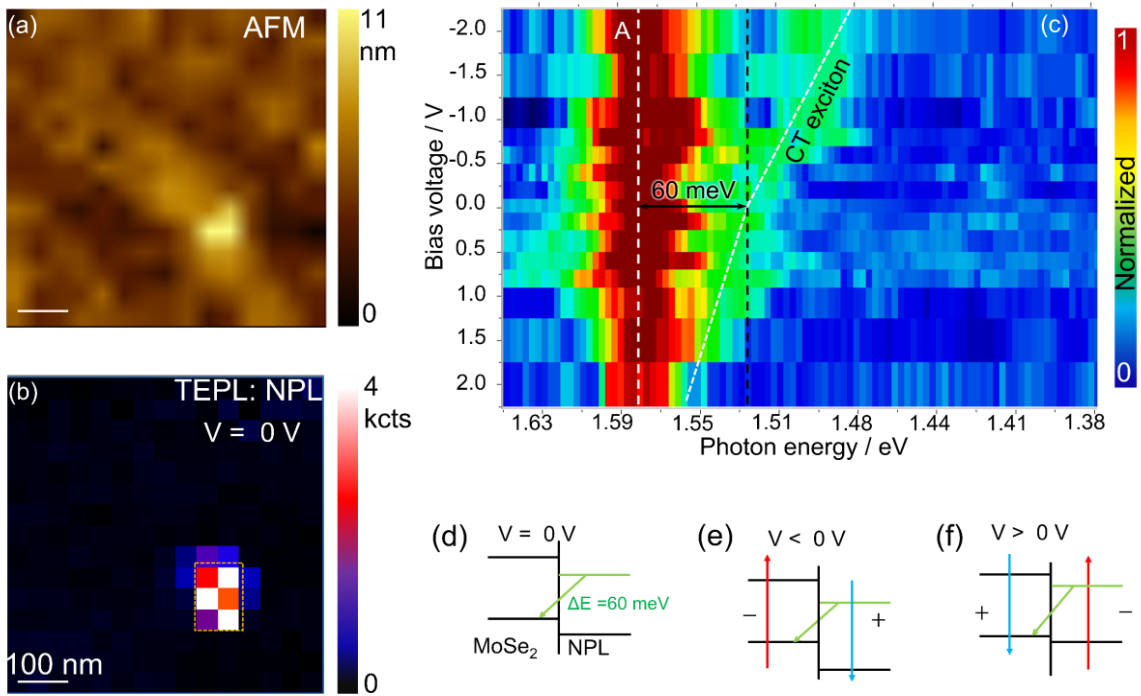

**Figure 3. *E*-field dependent tuning of CT excitons in MDH.** (a) TEPL map of NPL and (b) corresponding AFM topography of the NPL/MoSe$_2$ system acquired simultaneously at 0V bias voltage. (c) Contour plot of ***E***-field dependent TEPL spectra of the MDH system averaged over the rectangular area shown in (a). (d) – (f) Band alignment of the MDH at three different bias conditions.

NPLs exhibit strong (negligible) thickness dependent conduction band (CB) (valence band (VB)) tunability due to the small (large) conduction (valence) band offsets[26]. Hence, we can adjust the band alignment of the studied MDH systems via shell thickness to tune the CT exciton energy position. Fig. 2d displays NPL shell thickness dependent TEPL spectra of NPL/MoSe$_2$ MDH systems. As the shell thickness increases (from 4.5 ML to 7.5 ML), the CT exciton energy can be tuned up to 90 meV via tuning the band alignment. The process of band alignment tuning via NPL shell thicknesses is schematically presented in Fig. 2e-g. Larger shell thickness results in a smaller NPL bandgap, which moves the NPL CB minimum away (towards lower energy) from the MoSe$_2$CB minimum. This results in a larger band offset and consequently smaller CT exciton energies for thicker NPLs. In order to decouple CT excitons from strain and other local heterogeneity induced shift we also conducted a comprehensive TEPL investigation of the systems. Detailed studies of the local heterogeneities in the PL can be found in the supplementary information section IV.

CT excitons have an out-of-plane dipole moment similar to the case of interlayer exciton formation at a 2D/2D interface. This necessitated a study of the effect of out-of-plane *E*-field on the evolution of CT excitons in the 2D/QD MDHs. For the *E*-field dependent TEPL study, we used the same experimental configuration shown in Fig. 1a with a bias applied between the tip and the substrate during measurements. Fig. 3a,b presents a NPL TEPL map and corresponding AFM topography of the NPL/MoSe$_2$ MDH system acquired simultaneously at 0V bias. For each bias voltage a complete TEPL map was acquired for the same region of interest and an averaged TEPL spectra was created over the rectangle area marked in Fig. 3a. *E*-field dependent TEPL spectra of the MDH for the spectral region between 1.65 to 1.38 eV are shown in Fig. 3c. Evolution of NPL PL peak position as a function of bias voltage can be found in the supplementary information section V. For a NPL PL peak of 1.85 eV we observed the CT exciton at 60 meV below the MoSe$_2$ A exciton at 0V bias as schematically presented in the band diagram of Fig. 3d. As the bias increases in the negative direction, the CT exciton drifted further away in energy from the A exciton with a red shift of 120 meV observed at -2 V. The opposite trend was observed in the positive bias direction, though at a slower rate. It was not possible to decouple CT exciton from the A exciton peak above 1 V due to the close proximity and low single-to-noise ratio. Fig. 3d-f are sketched to explain the CT exciton evolution under an out-of-plane *E*-field. At 0 V, we have the standard band alignment for which the CT exciton is observed. However, at negative bias, the CB of MoSe$_2$ (NPL) increases (decreases) in energy. As a result, the CT exciton moves away energetically from the A exciton to a lower energy. The opposite situation occurs at a positive bias for which we observe the CT exciton move closer in energy to the A exciton of MoSe$_2$. A similar behavior was recently reported for delocalized analogous CT excitons (IL excitons) in a TMDC HB system[2,30].

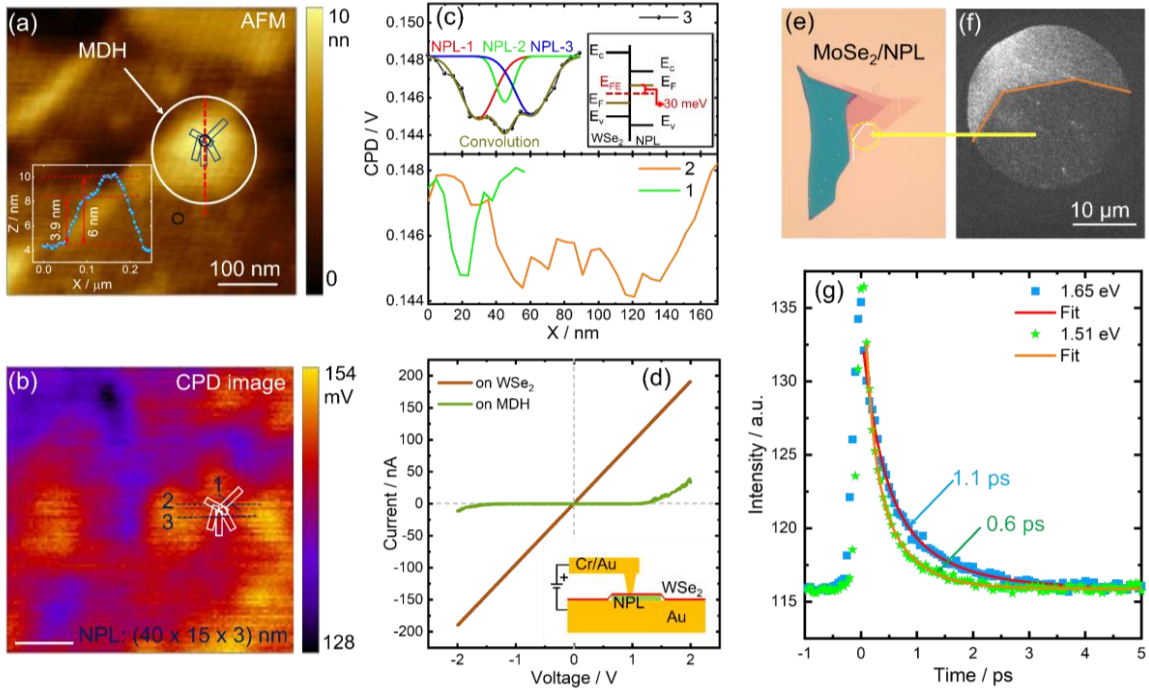

**Figure 4. Electrical and temporal characterization of MDH.** (a), (b) AFM and corresponding CPD images of a WSe$_2$/NPL MDH system. The height profile of the MDH along the dotted red line is shown in the inset of (a). (c) Three separate CPD profiles extracted along the dashed lines in (b). The top panel is fitted with three Gaussians. Inset: schematic illustration of MDH band alignment at charge equilibrium condition. (d) Locally measured I – V curves on two different spots marked by black circles in (a) using conductive AFM. (e), (f) Optical and PEEM images of a MDH sample used for tr-PEEM measurements. (g) Time-resolved PEEM of the MDH acquired using two different excitation laser energies to determine the CT exciton dynamics. Decay curves were spatially-averaged over the dashed circle in (e).

A further proof of forming type II band alignment (prerequisite for CT exciton formation) at NPL/TMDC (both MoSe$_2$ and WSe$_2$) heterointerfaces can be demonstrated by electrical characterization. Fig. 4a,b present AFM and corresponding contact potential difference (CPD) images of a NPL/WSe$_2$ MDH system. The height profile of the MDH (inset of Fig. 4a) indicates a cluster of NPLs with a possible pilling of two individual NPLs on top of each other since the thickness of each NPL should be around 3 nm (see sample 1 in Table S1). The CPD image in Fig. 4b exhibits a more interesting and informative electronic picture of the MDH. Even though ML-WSe$_2$ wraps the NPLs well and creates a single bulge in the topography, the CPD image shows traces of several NPL/WSe$_2$ interfaces. To extract the Fermi level information at charge neutrality conditions, we took three line profiles and plotted them in Fig. 4c. A gaussian fit to the line profile 1 provides a width of 13 nm (see supplementary information VI), which agrees well with the width of a single NPL. Interestingly, CPD line profiles 2 and 3 show several dips of the same value as line profile 1. Line profile 3 was fitted with three Gaussians: among them, the two outer ones have slightly higher width, and the middle one has the same width as line profile 1. Using the dimensions of the NPL, CPD line profiles and TEPL map (see Fig. SI-6), we sketched the MDH area with five NPLs, as shown in Fig. 4a,b. To explain the interfacial charge transfer phenomenon a schematic of the band diagram is plotted in the inset of Fig. 4c. Due to the Fermi level adjustment via interfacial charge transfer, the surface potential of WSe$_2$ decreases by 30 - 40 mV at the MDH interface, which is equivalent to a ~30 meV Fermi level rise at charge equilibrium conditions.

Fig. 4d displays the I – V curves measured locally using conductive-AFM on two different spots of the sample marked by black circles in Fig. 4a. More I – V curves are presented in the supplementary information section VI. Fig. 4d clearly demonstrates a rectification behavior for the MDH. Whereas, on WSe$_2$ we observed a linear I-V response due to direct electrical tunnelling or conduction through the ultra-thin layer of monolayer WSe$_2$.

Finally, we have used time-resolved photoemission electron microscopy (tr-PEEM) to investigate the exciton dynamics of CT excitons in MDH. Fig. 4e,f show the optical and PEEM image of the AOI region (outlined by circles). Exciton decay curves shown in Fig. 4g were derived from spatially-averaged dynamics within the AOI region. The femtosecond pump- probe tr-PEEM results of the MDH were acquired at two different pump excitation lasers: one at 1.65 eV covering the MoSe$_2$ exciton region and the other at 1.51 eV which predominately excites CT exciton. To fit the dynamics of the excitons both decay curves were fitted with bi-exponential functions with the fit parameters listed in the table S2. From the fit, the MoSe$_2$ A exciton lifetime (excited by 1.65 eV) is determined to be 1.1 ps. However, CT excitons generated at 1.51 eV show a shorter lifetime of 0.6 ps, which is in stark contrast to the lifetime of analogous ILXs in all 2D vdW heterostructures[2]. Indeed, the decay signal is the combination of both radiative and non-radiative recombination of excitons. At elevated temperature (tr-PEEM measurements were performed at RT) non-radiative decay through Augur scattering or charge trapping at defects dominates, which occurs on a faster time scale than radiative recombination [31,32]. Since $CdSe/Cd_xSZn_{1-x}S$ core/shell nanocrystals are known for surface defects/trap states[33], the probability of non-radiative recombination of CT excitons is higher at the $MoSe_2$/NPL interface than for the $MoSe_2$ A exciton in the 2D plane. A similar behavior was also observed by Bouleshba et al[34] at a $WS_2$/QD heterointerface. Hence, we observe a shorter CT exciton life time relative to the $MoSe_2$ A exciton in the present work.

**Conclusion**

In summary, we report the formation of CT excitons in a MDH containing 2D TMDCs and quasi-2D NPLs. We adopted TEPL to spatially resolve CT exciton formation sites at the single NPL/TMDC interface. To the best of our knowledge, this is the primary observation of such localized CT excitonic phenomena in a MDH at RT. To confirm the origin of CT excitons, we adopted a systematic approach via changing both the TMDC material and the shell thickness of NPLs. Both approaches give a wide range of tunability of the CT exciton energy, up to 100 meV. Out-of-plane ***E***-field dependent TEPL also provides an excellent knob to tune the CT exciton with a tunable range of 120 meV. Our work opens a new pathway for manipulating excited states at mixed-dimensional interfaces, which offers great promise both for fundamental studies and optoelectronic applications, opening doors to electrical control of colloidal quantum materials via electronic heterointerfaces with 2D materials.

**Methods**

**Synthesis of rectangular CdSe nanoplatelets:** The cadmium myristate precursor is prepared by following the literature[35]. Colloidal, rectangular CdSe nanoplatelets with a thickness of 4.5 monolayers are synthesized following the literature[36] with slight modifications[37,38] that are described in detail in the supplementary information section SI-I.

**Synthesis of square-like CdSe nanoplatelets:** Colloidal, square-like CdSe nanoplatelets with a thickness of 4.5 monolayers are synthesized by following prior literature[39] with slight modifications that are described in detail in the supplementary information section SI-I.

**Growth of $Cd_xZnS_{1-x}S$ shell:** Cadmium ($Cd(Ol)_2$) and zinc oleate ($Zn(Ol)_2$) are synthesized according to the literature[39,40]. The growth of a $Cd_xZnS_{1-x}S$ shell with increasing thickness on CdSe nanoplatelets is performed by following the literature[39] with minor modifications that are described in detail in the supplementary information section SI-I.

**MDH device preparation**: Both ultra-smooth Au and 5 nm $Al_2O_3$ coated Au substrates were used for the sample preparation. $Al_2O_3$ films were deposited on Au using ALD (Cambridge Nanotech) via chemical reaction of metal organic precursor, Trimethylaluminium with water vapors in each cycle at 150 °C, which typically yielded a deposition rate of 0.9 Å/cycle. To prepare the ultra-smooth Au surface, ~100 nm of Au was first deposited on Si with native oxide (without any adhesion layer) using the thermal evaporation technique. After that, the buried ultra-smooth Au face was stripped-off using an epoxy resin supported Si substrate, which were then used as the active substrates for the MDH devices.

To prepare the devices, a very dilute solution of NPL (0.001 mg/mL) was first spin coated at a speed of 3000 rpm for 60 s on the Au (or $Al_2O_3$/Au) substrate. After that, monolayer TMDCs were transferred on top of NPLs/Au via deterministic dry transfer method. All the sample were then annealed in an Ar/H2 atmosphere for 2 h at 120 °C.

**Micro-PL characterization**: A Horiba LabRam HR evolution equipped with 633 nm excitation laser and an electron multiplying charge coupled detector was used for micro-PL measurements. PL measurements were carried out using 100 l/mm grating and 100 µW power (measured at the sample surface) focused on the sample surface via a 100 x, 0.9 NA objective. PL maps were acquired using 500 x 500 nm step size, which is well above the diffraction limit at this wavelength. Signal acquisition time was set at 0.2 s during the mapping.

**TEPL characterization**: TEPL measurements were carried out using a Horiba NanoRaman platform consisting of the same Horiba LabRam HR evolution coupled with an AIST-NT AFM. Commercially available Au TERS tips were used for the measurements under ***p*-**polarized 633 nm excitation in side illumination/collection geometry at an angle of 65° from the normal to the sample surface. The laser power was kept at 20 µW focused onto the tip apex using a 100x 0.7 NA objective and the acquisition time was set at 0.2 s. Both the near-field PL spectra and far-field background were collected for each pixel one after another during TEPL map acquisition via a hybrid tip operating mode. Far-field background was measured during the tip normal oscillation period in non-contact mode and the near-field signal was collected by bringing the tip in contact to the sample. Both measuring steps were repeated one after another at each pixel to collect complete near-field plus far-field and far-field only PL maps of the sample. A step size of 20 x 20 nm was used for the all the TEPL mapping except for the ***E***-field dependent study, for which a 40 x 40 nm step size was used. For the ***E***-field dependent study, a bias was applied to the tip and kept constant during map acquisition.

**I – V characterization**: I – V characterization of the MDH were performed using conductive AFM and commercially available Cr/Au probes.

**Time-resolved PEEM**: The PEEM experiments were performed using a commercial titanium-sapphire oscillator (Griffin-10, KM Labs) producing sub-20 fs pulses centered at 780 nm at a 90 MHz repetition rate. A slit within the laser cavity is used for fundamental wavelength (740-840 nm range) and bandwidth tuning. Approximately 30% of the pulse is split to produce the second harmonic probe pulse in a 200 µm thick BBO crystal. The resulting blue light is recompressed in a $CaF_2$ prism pair to yield second harmonic pulses of less than 50 fs. A variable delay line controls the relative timing between the red (~800 nm) pump and blue (~400 nm) probe pulses. *P*-polarized laser pulses are recombined on a dichroic beam splitter and directed collinearly onto the PEEM sample at a 75° angle of incidence with respect to the surface normal. The spot sizes of the separate beams are adjusted such that typically the red pulse spot size is roughly 50% smaller than the blue pulse spot size at the sample position. A typical spot size for the red laser is 40 x 120 microns at the sample. PEEM images are collected as a function of probe delay time, yielding time resolved movies of photoelectron emission dynamics. Cross-correlation of the red and blue pulses yields time resolution of less than 80 fs for all wavelengths pairs used in this study. Typically, 50-100 mW of ~800 nm power is used in combination with 3-6 mW of ~400 nm laser light for experiments described herein.

**Author Contribution**

**Acknowledgements**

D.J. acknowledges primary support for this work by the Air Force Office of the Scientific Research (AFOSR) FA2386-20-1-4074. M.R. acknowledges support from Deutsche Forschungsgemeinschaft (DFG, German Research Foundation) for Walter Benjamin Fellowship (award no. RA 3646/1-1). The sample fabrication, assembly and characterization were carried out at the Singh Center for Nanotechnology at the University of Pennsylvania which is supported by the National Science Foundation (NSF) National Nanotechnology Coordinated Infrastructure Program grant NNCI-1542153. E.M. acknowledges support provided by the National Science Foundation under Grant No. DMR-2019444 (EM, CBM). S.S. acknowledges support from Basic

Science Research Program through the National Research Foundation of Korea (NRF) funded by the Ministry of Education (Grant No. 2021R1A6A3A14038492). Work by A.G.J and P.Z.K was supported by the U.S. Department of Energy (DOE), Office of Science, Basic Energy Sciences (BES). Work by B.T.O-C. was supported by the U.S. Department of Energy, (DOE), Office of Science, Biological and Environmental Research (BER). G.K. acknowledge support for this work by the Air Force Office of Scientific Research (AFOSR) FA2386-20-1-4074

**Data Availability**

The data that support the conclusions of this study are available from the corresponding author on request.

**References**

1. Alexandrou, A. *et al.* Electric-field effects on exciton lifetimes in symmetric coupled GaAs/Al0.3Ga0.7As double quantum wells. *Phys. Rev. B* **42**, 9225 (1990).

2. Rivera, P. *et al.* Observation of long-lived interlayer excitons in monolayer MoSe2–WSe2 heterostructures. *Nat. Commun. 2015 61* **6**, 1–6 (2015).

3. Wang, Z. *et al.* Evidence of high-temperature exciton condensation in two-dimensional atomic double layers. *Nat. 2019 5747776* **574**, 76–80 (2019).

4. Paik, E. Y. *et al.* Interlayer exciton laser of extended spatial coherence in atomically thin heterostructures. *Nat. 2019 5767785* **576**, 80–84 (2019).

5. Islam, M. N. *et al.* Electroabsorption in GaAs/AlGaAs coupled quantum well waveguides. *Appl. Phys. Lett.* **50**, 1098 (1998).

6. Steiner, T. W., Wolford, D. J., Kuech, T. F. & Jaros, M. Auger decay of X-point excitons in a type II GaAs/AlGaAs superlattice. *Superlattices Microstruct.* **4**, 227–232 (1988).

7. Jiang, Y., Chen, S., Zheng, W., Zheng, B. & Pan, A. Interlayer exciton formation,

relaxation, and transport in TMD van der Waals heterostructures. *Light Sci. Appl.* **10**, 1–29 (2021).

8. Rivera, P. *et al.* Interlayer valley excitons in heterobilayers of transition metal dichalcogenides. *Nat. Nanotechnol.* **13**, 1004–1015 (2018).

9. Mak, K. F. & Shan, J. Opportunities and challenges of interlayer exciton control and manipulation. *Nat. Nanotechnol.* **13**, 974–976 (2018).

10. Ruiz-Tijerina, D. A. & Fal'Ko, V. I. Interlayer hybridization and moiré superlattice minibands for electrons and excitons in heterobilayers of transition-metal dichalcogenides. *Phys. Rev. B* **99**, 30–32 (2019).

11. Brem, S., Linderälv, C., Erhart, P. & Malic, E. Tunable Phases of Moiré Excitons in van der Waals Heterostructures. *Nano Lett.* **20**, 8534–8540 (2020).

12. Interlayer excitons and how to trap them. *Nat. Mater. 2020 196* **19**, 579–579 (2020).

13. Yu, H., Liu, G. Bin, Tang, J., Xu, X. & Yao, W. Moiré excitons: From programmable quantum emitter arrays to spin-orbit–coupled artificial lattices. *Sci. Adv.* **3**, 1–8 (2017).

14. Seyler, K. L. *et al.* Signatures of moiré-trapped valley excitons in MoSe2/WSe2 heterobilayers. *Nature* **567**, 66–70 (2019).

15. Li, W., Lu, X., Dubey, S., Devenica, L. & Srivastava, A. Dipolar interactions between localized interlayer excitons in van der Waals heterostructures. *Nat. Mater.* **19**, 624–629 (2020).

16. Jadwiszczak, J. *et al.* Mixed-Dimensional 1D/2D van der Waals Heterojunction Diodes and Transistors in the Atomic Limit. *ACS Nano* **16**, 1639–1648 (2022).

17. Mouafo, L. D. N. *et al.* 0D/2D Heterostructures Vertical Single Electron Transistor. *Adv. Funct. Mater.* **31**, 2008255 (2021).

18. Zereshki, P. *et al.* Observation of charge transfer in mixed-dimensional heterostructures formed by transition metal dichalcogenide monolayers and PbS quantum dots. *Phys. Rev. B* **100**, 1–8 (2019).

19. Xiao, J. *et al.* Type-II Interface Band Alignment in the vdW PbI2-MoSe2Heterostructure. *ACS Appl. Mater. Interfaces* **12**, 32099–32105 (2020).

20. Jariwala, D., Marks, T. J. & Hersam, M. C. Mixed-dimensional van der Waals heterostructures. *Nat. Mater.* **16**, 170–181 (2017).

21. Zhu, X. *et al.* Charge Transfer Excitons at van der Waals Interfaces. *J. Am. Chem. Soc.* **137**, 8313–8320 (2015).

22. Goodman, A. J., Dahod, N. S. & Tisdale, W. A. Ultrafast Charge Transfer at a Quantum Dot/2D Materials Interface Probed by Second Harmonic Generation. *J. Phys. Chem. Lett.* **9**, 4227–4232 (2018).

23. Zhang, K. *et al.* Electrical control of spatial resolution in mixed-dimensional heterostructured photodetectors. *Proc. Natl. Acad. Sci. U. S. A.* **116**, 6586–6593 (2019).

24. Burdanova, M. G. *et al.* Intertube Excitonic Coupling in Nanotube Van der Waals Heterostructures. *Adv. Funct. Mater.* **32**, 2104969 (2022).

25. Ithurria, S. *et al.* Colloidal nanoplatelets with two-dimensional electronic structure. *Nat. Mater. 2011 1012* **10**, 936–941 (2011).

26. Eshet, H., Grünwald, M. & Rabani, E. The electronic structure of CdSe/CdS Core/shell seeded nanorods: Type-I or quasi-type-II? *Nano Lett.* **13**, 5880–5885 (2013).

27. Gong, C. *et al.* Band alignment of two-dimensional transition metal dichalcogenides: Application in tunnel field effect transistors. *Appl. Phys. Lett.* **103**, 053513 (2013).

28. Rahaman, M. *et al.* Observation of Room-Temperature Dark Exciton Emission in

Nanopatch-Decorated Monolayer WSe2 on Metal Substrate. *Adv. Opt. Mater.* (2021) doi:10.1002/adom.202101801.

29. Park, K. D., Jiang, T., Clark, G., Xu, X. & Raschke, M. B. Radiative control of dark excitons at room temperature by nano-optical antenna-tip Purcell effect. *Nat. Nanotechnol.* **13**, 59–64 (2018).

30. Ciarrocchi, A. *et al.* Polarization switching and electrical control of interlayer excitons in two-dimensional van der Waals heterostructures. *Nat. Photonics* **13**, 131–136 (2019).

31. Javaux, C. *et al.* Thermal activation of non-radiative Auger recombination in charged colloidal nanocrystals. *Nat. Nanotechnol. 2013 83* **8**, 206–212 (2013).

32. Moody, G., Schaibley, J. & Xu, X. Exciton dynamics in monolayer transition metal dichalcogenides [Invited]. *JOSA B, Vol. 33, Issue 7, pp. C39-C49* **33**, C39–C49 (2016).

33. Minotto, A. *et al.* Role of core-shell interfaces on exciton recombination in CdSe-CdxZn1-xS quantum dots. *J. Phys. Chem. C* **118**, 24117–24126 (2014).

34. Boulesbaa, A. *et al.* Ultrafast Charge Transfer and Hybrid Exciton Formation in 2D/0D Heterostructures. *J. Am. Chem. Soc.* **138**, 14713–14719 (2016).

35. Carion, O., Mahler, B., Pons, T. & Dubertret, B. Synthesis, encapsulation, purification and coupling of single quantum dots in phospholipid micelles for their use in cellular and in vivo imaging. *Nat. Protoc. 2007 210* **2**, 2383–2390 (2007).

36. She, C. *et al.* Red, Yellow, Green, and Blue Amplified Spontaneous Emission and Lasing Using Colloidal CdSe Nanoplatelets. *ACS Nano* **9**, 9475–9485 (2015).

37. Marino, E. Assembling nanocrystal superstructures. (Universiteit van Amsterdam, 2019).

38. Marino, E. *et al.* Repairing Nanoparticle Surface Defects. *Angew. Chemie Int. Ed.* **56**, 13795–13799 (2017).

39. Rossinelli, A. A. *et al.* Compositional Grading for Efficient and Narrowband Emission in CdSe-Based Core/Shell Nanoplatelets. *Chem. Mater.* **31**, 9567–9578 (2019).

40. Hendricks, M. P., Campos, M. P., Cleveland, G. T., Plante, I. J. La & Owen, J. S. A Tunable library of substituted thiourea precursors to metal sulfide nanocrystals. *Science (80-. ).* **348**, 1226–1230 (2015).

**Supplementary information:**

# Tunable Localized Charge Transfer Excitons in a Mixed Dimensional van der Waals Heterostructure

Mahfujur Rahaman, Emanuele Marino, Alan G. Joly , Seunguk Song, Zhiqiao Jiang, Brian T. O'Callahan, Daniel J. Rosen, Kiyoung Jo, Gwangwoo Kim, Patrick Z. El-Khoury, Christopher Murrey, and Deep Jariwala

**Table of Content:**

## I. Synthesis and characterization of CdSe/Cd$_x$SZn$_{1-x}$S NPLs

**Synthesis of rectangular CdSe nanoplatelets**: Cadmium myristate precursor is prepared by following the literature[1]. Colloidal, rectangular CdSe nanoplatelets with a thickness of 4.5 monolayers are synthesized following the literature[2] with slight modifications[3,4].

340 mg of finely-ground cadmium myristate and 28 mL 1-octadecene (technical grade, ODE) are added to a 100 mL three-necked round-bottom flask with a 1-inch octagonal stir bar. The central neck is connected to the Schlenk line through a 100 mL bump trap, one of the side necks is equipped with a thermocouple adapter and thermocouple, and the other one is fitted with a rubber stopper. With a heating mantle, the flask is degassed at 100 °C for 30 min. In the meanwhile, a dispersion of 0.15 M selenium in ODE is prepared by sonication for at least 20 min. After switching to the atmosphere of the flask to nitrogen, the temperature of the reaction is increased to 220°C. 2 mL of 0.15 M Se/ODE dispersion is quickly injected by using a 22 mL plastic syringe equipped with a 16 G needle. After 20 seconds, 120 mg of finely-group cadmium acetate is added to the flask by temporarily removing the stopper. The flask is carefully rocked to ensure that the cadmium acetate powder does not stick to the side walls of the flask. The reaction is kept at 220°C for 14 minutes and then rapidly cooled with a water bath. 12 mL of oleic acid (technical grade, OA) and 22 mL of hexane are added when the temperature reaches 160°C and 70°C, respectively.

**Synthesis of square-like CdSe nanoplatelets**: Colloidal, square-like CdSe nanoplatelets with a thickness of 4.5 monolayers are synthesized by following the literature[5] with slight modifications.

680 mg of finely-ground cadmium myristate, 48 mg of selenium, and 60 mL of ODE are added to a 250 mL three-necked round-bottom flask with a 1-inch octagonal stir bar. The central neck is connected to the Schlenk line through a 100 mL bump trap, one of the side necks is equipped with a thermocouple adapter and thermocouple, and the other one is fitted with a rubber stopper. The contents of the flask are degassed at room temperature for 30 min. The atmosphere of the flask is then switched to nitrogen, and the temperature is raised to 240 °C using a heating mantle. When the temperature reaches 200 °C, 256 mg of finely-ground cadmium acetate dihydrate is added by temporarily removing the rubber stopper. The flask is carefully rocked to ensure that the cadmium acetate powder does not stick to the side walls of the flask. The mixture is kept at 240 °C for 8−9 min. Specifically, 8 min results in an average nanoplatelet area of 11.4 nm x 10.1 nm, while 9 min results in an average area of 17.8 nm x 15.4 nm. Afterward, the reaction flask is cooled using an air gun. At 185 °C, 3 mL of OA are added. At 130 °C, the flask is lowered in a water bath and cooled to room temperature.

**Washing procedure of CdSe nanoplatelets**: The nanoplatelets are washed by following a procedure reported in the literature with modifications[5]. The mixture is first centrifuged at 8586 g for 10 min. The precipitate is then redispersed in 10 mL of hexane. The suspension is left undisturbed for 1 h, and then centrifuged at 6574 g for 7 min. The precipitate is discarded as it contains undesired 3.5 monolayer nanoplatelets. The supernatant is retained and transferred to a new centrifuge tube. 10 mL of methyl acetate are added to the supernatant, followed by centrifugation at 5668 g for 10 min. 6 mL hexane is used to redisperse the precipitate. Measuring the optical absorption spectrum is useful to confirm the removal of the unwanted 3.5 monolayer nanoplatelets, which are characterized by a lowest-energy absorption peak at 462 nm, while the 4.5 monolayer nanoplatelets are characterized by a lowest-energy absorption peak at 512 nm. If 3.5 monolayer nanoplatelets are still present in the dispersion, they can be removed by titrating methyl acetate and centrifuging until all 3.5 monolayer nanoplatelets are successfully removed. The final dispersion is stored in a glass vial in the dark.

**Growth of Cd$_x$ZnS$_{1-x}$S shell**: Cadmium (Cd(Ol)$_2$) and zinc oleate (Zn(Ol)$_2$) are synthesized according to the literature[5,6]. The growth of the Cd$_x$ZnS$_{1-x}$S shell on CdSe nanoplatelets is performed by following the literature[5] with minor modifications.

10 mL of ODE, 0.4 mL of OA, 90 mg of cadmium oleate, 167.5 mg of zinc oleate, and an amount of 4.5 monolayer CdSe nanoplatelets in hexane equivalent to a 1mL with an optical density of 120/cm at the lowest-energy absorption peak are added to a 100 mL three-necked round-bottom flask with a 1-inch octagonal stir bar. The central neck is connected to the Schlenk line through a 100 mL bump trap, one of the side necks is equipped with a thermocouple adapter and thermocouple, and the other one is fitted with a rubber stopper. The mixture is degassed for 35 min at room temperature and for 15 min at 80 °C. In the meanwhile, a solution of 83 µL of 1-octanethiol (OT) in 7 mL of degassed ODE and 2 mL of degassed OA is prepared in the glove box and loaded in a plastic syringe. 2 mL of degassed oleylamine are added to a second plastic syringe. The two syringes are removed from the glove box. Afterward, the atmosphere of the reaction flask is switched to nitrogen, and 2 mL of OAm are injected. Using a heating mantle, the temperature of the reaction flask is increased to 300 °C. At 165 °C, the solution of OT in ODE and OA is injected at a rate of 4.5 mL/h. After complete injection, the temperature of the reaction is maintained for an additional 40 min. The reaction mixture is cooled down to 240 °C by using an air gun, followed by using a water bath to cool to room temperature. At 40 °C, 5 mL of are added.

**Washing procedure of CdSe/Cd$_x$ZnS$_{1-x}$S nanoplatelets**: The reaction mixture is centrifuged at 6000 g for 6 min. The precipitate is redispersed in 5 mL of hexane while the supernatant is discarded. Methyl acetate is added to the dispersion until the mixture turned turbid, followed by centrifugation at 6000 g for 10 min. This process is repeated. The precipitate is redispersed in 3 mL of hexane and centrifuged at 6000 g for 7 min. The precipitate is discarded, containing aggregated nanoplatelets. The supernatant is retained and filtered through a 0.2 µm PVDF or PTFE syringe filter. The dispersion is stored in the dark under ambient conditions. The final dispersion is stored in a glass vial in the dark.

The thickness of the shell can be tuned by increasing the amounts of $Cd(Ol)_2$, $Zn(Ol)_2$, and OT while preserving the same ratios. Specifically, we use the different amounts of precursors to obtain varied thicknesses as measured by TEM, which are summarized in Table S1.

**Table S1. Synthesis parameters and physical dimension of CdSe/Cd$_x$SZn$_{1-x}$S NPL**

| Sample # | Cd(Ol)$_2$ (mg) | Zn(Ol)$_2$ (mg) | OT (µL) | Length (nm) | Width (nm) | Thickness (nm) |
|---|---|---|---|---|---|---|
| 1 | 90 | 167.5 | 83 | 40.2 ± 2.9 | 16.1 ± 1.7 | 2.8 ± 0.5 |
| 2 | 120 | 223.3 | 111 | 11.4 ± 1.3 | 10.1 ± 1.1 | 4.4 ± 0.5 |
| 3 | 150 | 279.2 | 138 | 17.8 ± 1.8 | 15.4 ± 1.8 | 5.6 ± 0.4 |

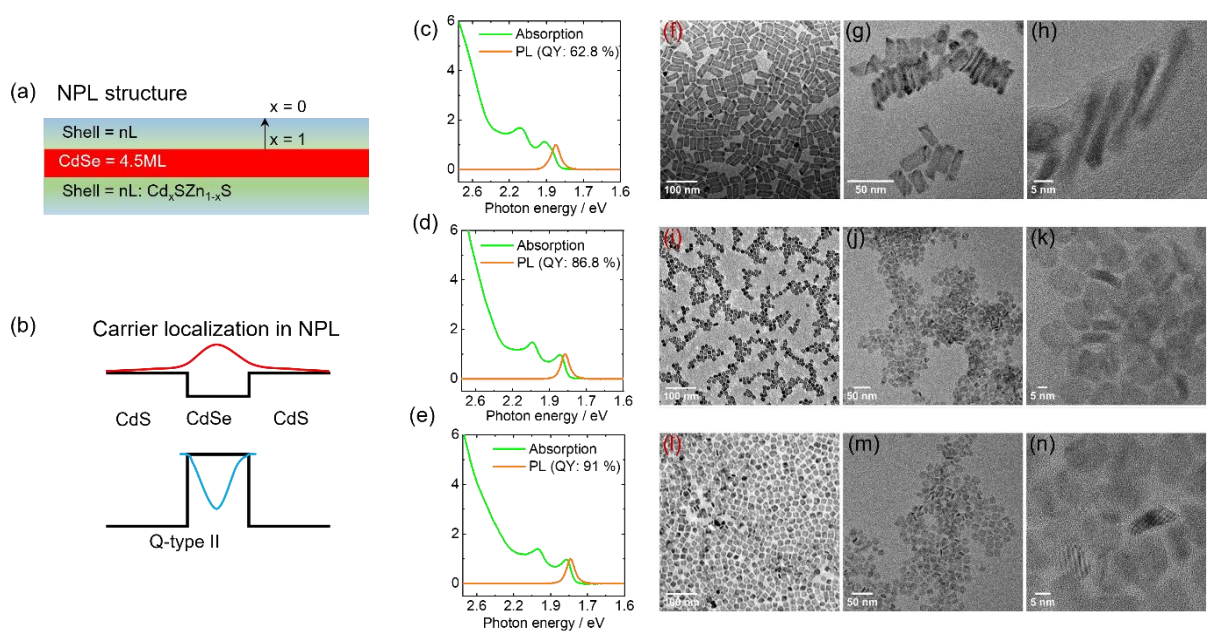

**Figure SI-1: Synthesis and characterization of NPLs.** (a) Schematic of core/shell CdSe/Cd$_x$SZn$_{1-x}$S NPL. Around the CdSe core, the shell comprised of CdS and further moving away from the core Cd is replaced by Zn (b) Quasi type II band structure of NPL. Due to the large band offset, holes are localized in the CdSe core. On the other hand, electrons are delocalized in the core/shell due to a small band offset. (c)–(e) Normalized absorption and PL spectra of three different samples of NPLs presented in Table S1, respectively. (f)–(h), (i)–(k), and (l)–(n) are the TEM images of the NPLs with optical properties shown in (c), (d), and (e), respectively.

**TEM imaging**: TEM was performed on a JEOL F200 multipurpose microscope operating at 200 kV.. To prepare the dispersed nanocrystals for imaging, we drop cast 10 µL of a dilute (~0.1 mg/mL) dispersion of nanocrystals in hexane on a lacey carbon-coated TEM grid (EMS). The grid was dried under 0.1 Torr vacuum for 1 hour prior to imaging to remove excess organic solvent.

**Spectrophotometry**: Absorption spectra of nanocrystal dispersions in toluene were measured by using a Cary 5000 UV-Vis-NIR spectrophotometer.

**PLQY**: PLQY measurements were performed by using the integrating sphere module of an Edinburgh FLS1000 Photoluminescence Spectrometer. The NCs were dispersed at a concentration corresponding to an absorbance of 0.1 at the excitation wavelength.

## II. Micro-PL and TEPL characterization of MDH

Fig. SI-2ia, b are the optical and PL intensity overlay images (NPL and $MoSe_2$) of the area of interest (AOI) region shown in the main text Fig. 1b. Two representative PL spectra of NPL and ML-$MoSe_2$ and their differential response acquired from two pixels in the PL map are presented in Fig. SI-2ic. Apart from the main excitonic peaks of NPL and $MoSe_2$, we can also observe the CT excitonic peak around 1.51 eV. AFM image of the AOI is shown in Fig. SI-2id. The height profile of the monolayer part of the flake along the red line can be seen in the inset, which indicates the thickness of the flake to be $1.2 \pm 0.5$ nm. The slightly higher thickness of the flake can be due to the organic absorbates and water molecules trapped beneath the flake.

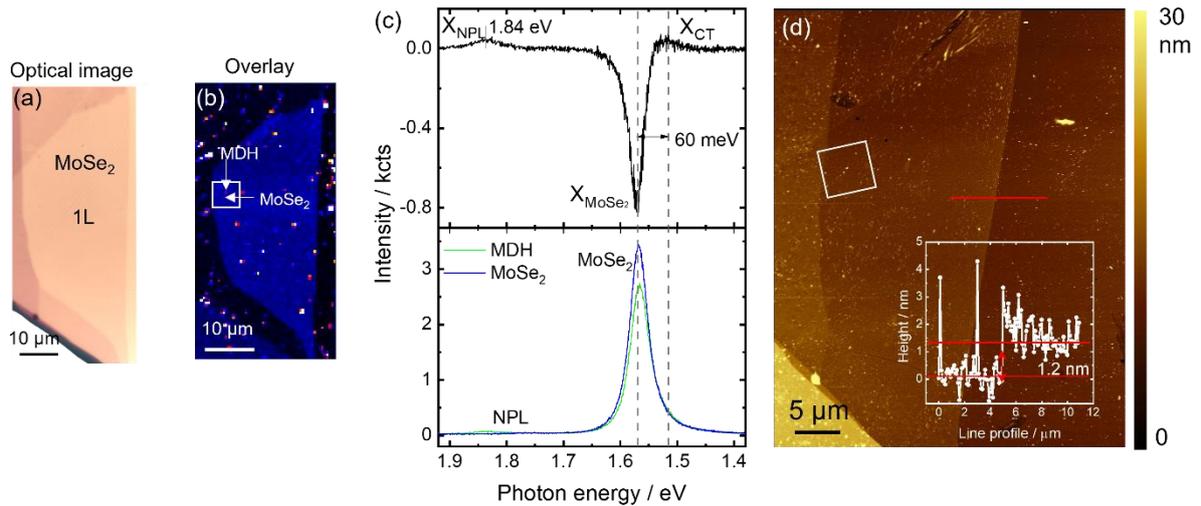

**Figure SI-2i. Micro-PL characterization of the MDH**. (a) Optical and (b) PL intensity overlay images of the area of interest (AOI) shown in Fig. 1b in the main text. (c) Two representative PL spectra of NPL and ML-MoSe$_2$ and their differential response acquired from two pixels in the PL map are presented in (b). (d) AFM image of the AOI. Inset is the height profile of the flake along the red line.

Fig. SI-2iia presents the magnified AFM image of the area marked by a white rectangle in both Fig. SI-2ib and d. Even though there are several bright spots that can be identified on the image, not all of them are from NPLs, as confirmed by both high-resolution micro-PL and TEPL mapping. One NPL/MoSe$_2$ interface is marked by the white rectangle on the image for which near-field (NF) data are presented here. Fig. SI-2iib is the AFM image of the TEPL map area marked by the rectangle in Fig. SI-2iia taken simultaneously with the TEPL map. At the center of the AFM image, the bright topography resembled the NPL/MoSe$_2$ interface with a height of 5 nm, which is the thickness of a single NPLs of sample 2 in Table S1. A step size of 20 x 20 nm was used for the TEPL measurements. The NF+FF (far-field) and the FF map of the NPL are presented in Fig. SI-2iic,d, respectively. The NF+FF map of NPL directly correlates with the AFM image in Fig. SI-2iib. However, the background subtracted NF map in Fig. SI-2iie resolves the NPL perfectly. The combination of the AFM image in Fig. SI-2iib, NF map in Fig. SI-2iie, and the dimension parameters for NPL presented in the second row of Table S1 clearly show that we have a single NPL at the centre.

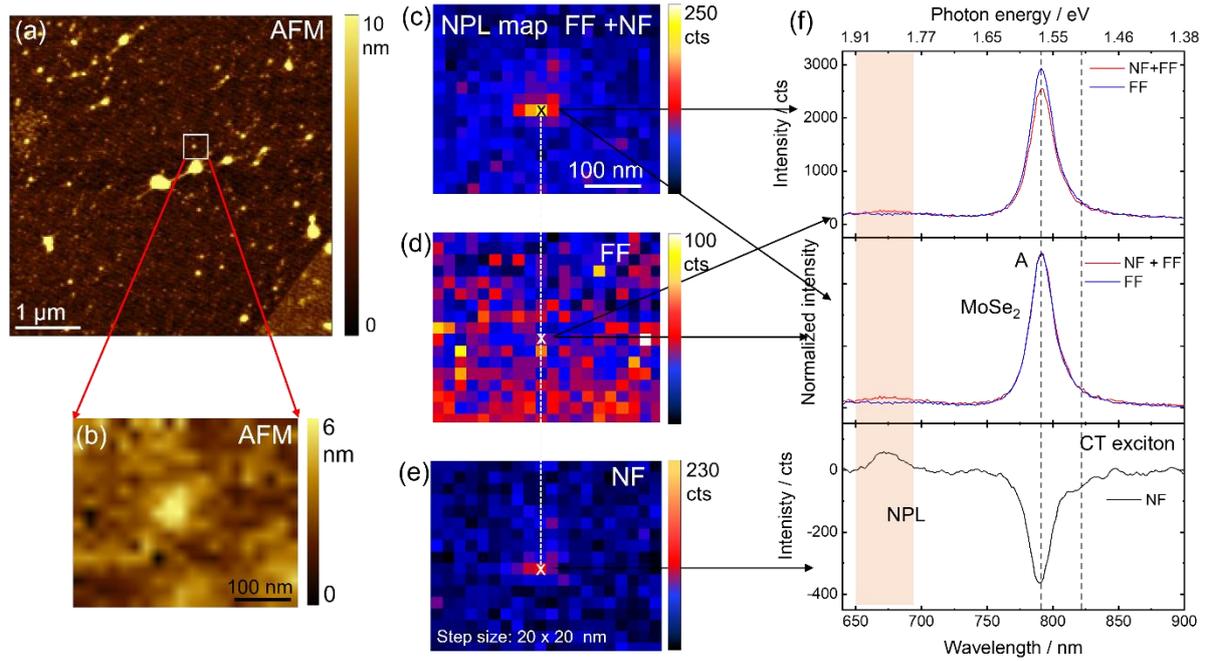

**Figure SI-2ii. TEPL of MDH**. (a) Magnified AFM image of the area marked by a rectangle in both SI-II-1b and d. (b) AFM topography of the TEPL area marked by a rectangle in (a) acquired simultaneously with TEPL. (c) FF+NF, (d) FF, and (e) FF background subtracted NF map of NPL. (f) Three corresponding FF+NF, FF, and NF PL spectra of single NPL/MoSe$_2$ heterointerface acquired from the marked pixel in (c) – (e) respectively. The orange shade indicates the spectral range for which NPL maps were created.

The corresponding NF+FF, FF, and NF spectra of the single NPL/MoSe$_2$ heterointerface are presented in Fig. SI-2iif. The PL peak of NPL is around 1.85 eV and agrees well Fig. SI-1d. The energy difference between MoSe$_2$ A exciton and the CT exciton is measured to be 60 meV for this NPL, which agrees well with the NPL/MoSe$_2$ system discussed in Fig. 3d of main manuscript. Note that, for this specific TEPL measurement, PL intensity around MoSe$_2$ spectral region in the NF+FF map was smaller than the FF map. This could be due to the combination of two factors. The more important one is the geometry reason which will be discussed in the next section. The second one is the physical mechanism of the formation of CT exciton. As explained in Fig. 3a in the main manuscript, photoexcited charge (electrons) transfer occurs from MoSe$_2$ to NPL, and then electrons are recombined radiatively with the holes in the valence band of MoSe$_2$. Consequently, CT excitons are formed, and the overall PL quantum yield of MoSe$_2$ reduces. Since this optical phenomenon occurs locally, FF PL is not sensitive to this. In comparison, PL intensity in the NF measurement decreases.

### III. Role of far-field background in near-field measurements

The major challenge of near-field measurements is the sensitivity of the plasmonic field; as such, the signal enhanced by a tip needs to overcome the FF background signal. The excitation laser used in the measurement always covers a far bigger area (~ µm$^2$) than the local optical site (~ nm$^2$), which is roughly $10^3 – 10^4$ times larger (see Fig. SI-3a). For a nano-object such as NPL in the current investigation, this can be less challenging since the size of the NPL is comparable to the spatial resolution of TEPL, which suppress the FF signal significantly. However, for 2D material, the large extent of the crystal along the x-y plane poses a significant challenge since the local enhancement needs to be larger than the area factor ($10^3 – 10^4$) to have a positive NF signal in Fig. SI-2iif. Since the CT exciton is close to the A exciton of MoSe$_2$, it is also affected by the local enhancement and the FF contribution of the A exciton in terms of relative signal enhancement. In our current investigation, we always obtained a positive NF signal for NPL. However, for MoSe$_2$ and CT excitons, we obtained positive NF signals, usually with brand new (and less unmodified) Au tips. Generally, with a used tip after several experimental sessions, the NF signal for the MoSe$_2$ excitonic spectral regime turns negative. This indicates that the plasmonic strength of the tip gets reduced after several sessions, which was also confirmed by the reflectance measurements conducted before each TEPL session.

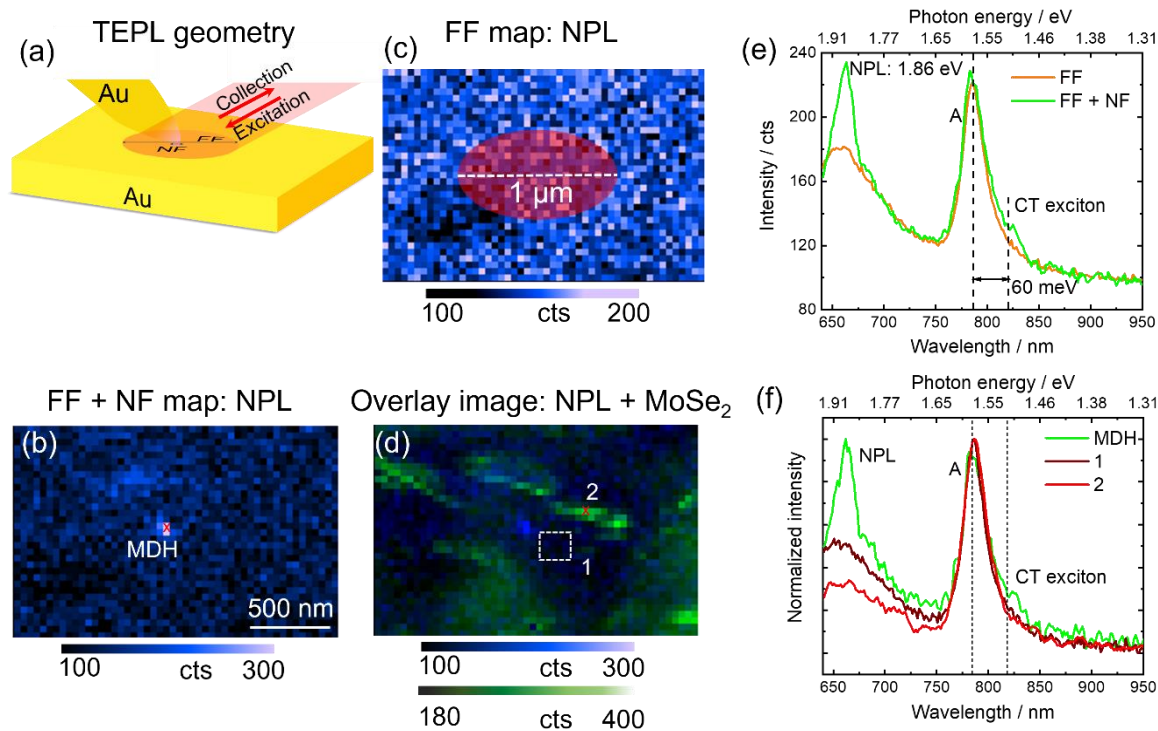

**Figure SI-3. Role of FF background in resolving CT excitons**. (a) Schematic diagram of TEPL geometry showing a proportionate FF and NF probing area. (b) FF+NF and (c) FF maps of NPL PL intensity acquired during TEPL measurements. (d) Overlay image of FF+NF maps of NPL and MoSe$_2$ PL intensity. (e) Two representative FF+NF and FF spectra acquired in a proportionate scale from a pixel and overaged over a ellipsoid, as indicated in the respective images in (b) and (c). (f) normalized FF+NF spectra of MDH is compared with two nearby MoSe2 spectra acquired from a pixel and averaged over a small area as marked in (d) respectively.

Fig. SI-3b, c show FF+NF and FF maps of NPL acquired from another sample and with a relatively unmodified Au tip. The FF+NF map clearly resolves the single NPL at the centre, which FF map failed to detect. What is interesting to see here is that even though the NF+FF spectrum clearly creates a stark contrast for the NPL PL peak compared to the FF one, the FF MoSe$_2$ PL signal is almost equal to the FF+NF signal. As a comparison, an overlay image of the NPL and MoSe$_2$ FF+NF map is created and presented in Fig. SI-3d. Two comparative TEPL spectra were acquired from the NF+FF and FF map of NPL with a proportionate area, as shown in Fig. SI-3e. In addition to the two excitonic peaks of NPL and MoSe$_2$, we can clearly identify the CT exciton as a shoulder at the lower energy end of MoSe$_2$, roughly 60 meV away from the A exciton in the FF+NF spectra. To compare with the excitonic behaviour of MoSe$_2$

nearby, we acquired two additional spectra, one acquired from a single pixel and the other averaged over a small area as indicated in the overlay image and plotted together with the NPL spectra in Fig. SI-3f. As can be seen, none of the $MoSe_2$ spectra nearby exhibits a similar shoulder at the lower energy end, confirming that it is an interfacial exciton, namely a CT exciton.

## IV. Decoupling CT excitons from local heterogeneities

One of the consequences of annealing 2D materials is that degassing creates nano-bubbles, which become the source of localized strain on the material. Ultra-thin TMDCs are very sensitive to localized strain, as reported previously[7]. Therefore, to decouple the CT excitons from the localized strain-induced spectral modification, we conducted a thorough TEPL investigation. Fig. SI-4a shows an AFM image containing heterogeneous morphological information. With the help of NF PL intensity maps (overlay image in Fig. SI-4b) and PL peak position map of $MoSe_2$ (Fig. SI-4c) we can decouple all the heterogeneous morphological components. For example, the biggest patches for which we observed the most enhanced PL and no (or small) peak shift are made of PDMS residuals forming on top of $MoSe_2$. Since the ML-TMDCs were transferred via the PDMS-assisted dry transfer method, annealing of such samples can create PDMS nano-patches. Similar TEPL behaviour was also observed recently[8]. The small red shift of the PL peak can be due to the local dielectric screening effect. Next in line are the nano-bubbles, which are easily identifiable via Fig. SI-4c. Since nano-bubbles are responsible for local strain, spectral modification, as well as red shifting, should also be the strongest in these areas. Three of these nano-bubbles are marked with red arrows in the AFM image. Finally, MDH interfaces can be identified from the overlay image of both NPL and $MoSe_2$ PL intensity maps in Fig. SI-4b. Six representative TEPL spectra of $MoSe_2$ on a flat substrate from nano-bubbles and nano-patches are plotted in Fig. SI-4d for comparison. As expected, PL spectra from nano-bubble areas are the most red-shifted due to localized strain. To compare the local deformation of MoSe$_2$ film caused by nano-bubbles and NPLs, we acquired height profiles of MoSe$_2$ along the three dashed red lines in the AFM image and plotted in Fig. SI-4e. One can see that NPL-induced local bending of MoSe$_2$ is the lowest of all, with the smallest aspect ratio (height-to-length ratio). Therefore, if there is any strain induced spectral modification due to NPL morphology, then it should be the minimum (negligible) on NPLs. Interestingly, as can be seen in the TEPL spectra of MoSe$_2$ on nano-bubbles and NPL presented in Fig. SI-4f energy offset for CT exciton is comparable to the strain induced shift caused by nano-bubbles, and therefore cannot be explained by localized strain.

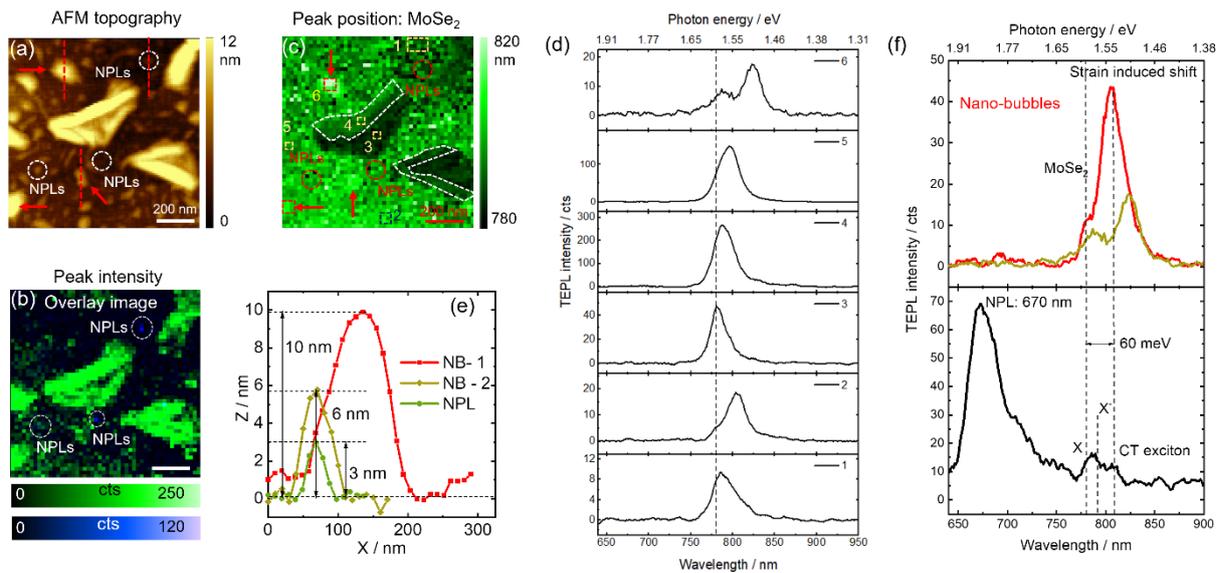

**Figure SI-4. Decoupling CT excitons from local heterogeneities**. (a) AFM image of a NPL/MoSe$_2$ sample containing heterogeneous information. (b) Corresponding NF PL intensity overlay image of MoSe$_2$ and NPL. (c) Corresponding NF PL peak position map of MoSe$_2$. (d) Representative TEPL spectra of MoSe$_2$ acquired from different regions as marked in the peak position map in (c). (e) Local bending profiles of MoSe$_2$ on top of nano-bubbles and NPL. (f) Comparison of TEPL spectra between MDH and MoSe$_2$ on nano-bubbles.

## V.  E-field dependent quantum confined stark effect in NPL

Fig. SI-5a,b are the AFM and the corresponding TEPL map of NPL acquired simultaneously at 0 V bias. As discussed in the main manuscript in Fig. 3a-c, a TEPL map was acquired at each bias voltage. An averaged TEPL spectra within the NPL spectra region (1.94 – 1.68 eV) was acquired at each bias and presented in Fig. SI-5c. To understand the quantum confined stark effect (QCSE) better, we plotted the evolution of the peak position of the NPL as a function of bias voltage in Fig. SI-5d. The PL peak position shows a dip at a bias voltage of 0.8 V. We studied QCSE effect in several NPLs. However, we did not observe a similar QCSE trend for the individual NPLs studied in this work. This could be due to the fluctuation of the local electric field in and around the individual NPLs originated from several effects such as photoionization, screening due to external charges, trap states, etc., as discussed in the previous literature[9].

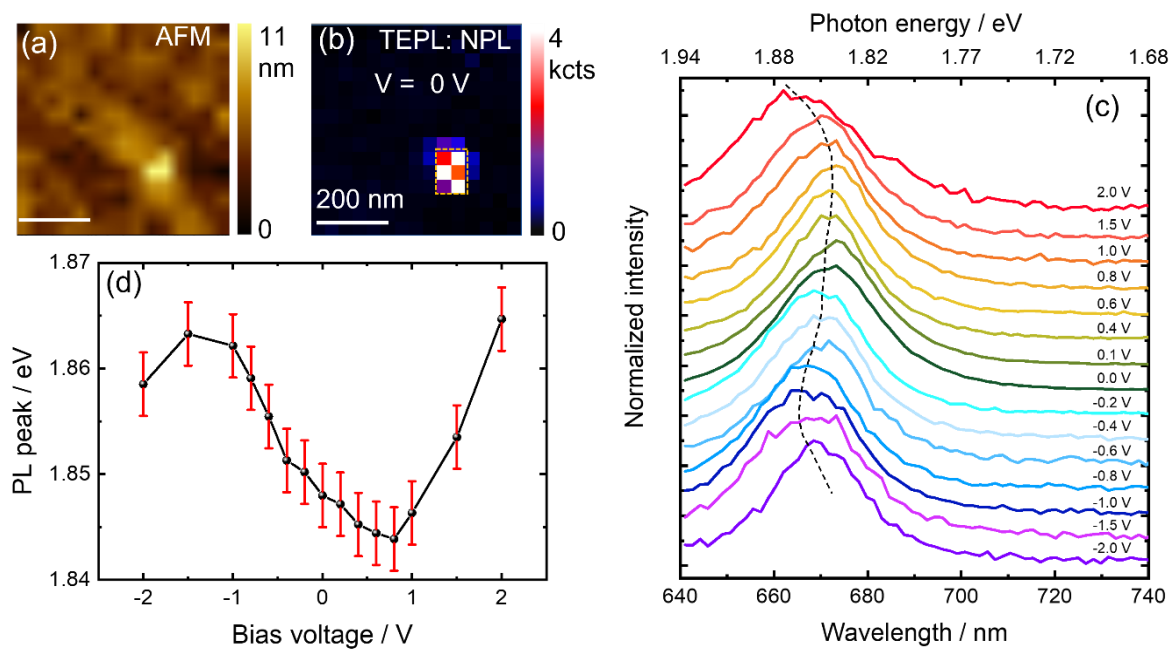

**Figure SI-5. *E*-field dependent QCSE effect in NPL**. (a) AFM topography and (b) corresponding NPL TEPL map at 0V bias acquired simultaneously. (c) Averaged TEPL spectra of NPL at different bias voltage. All the spectra were averaged over six pixels as marked by a dashed rectangle in (b). (d) Evolution of peak position of the PL peak shown in (c) as a function of bias voltage.

## VI. Local electrical characterization of MDH

An AFM image and its corresponding TEPL NPL map of MDH consisting of NPL/WSe$_2$ on bare Au is presented in Fig. SI-6a,b. Two representative TEPL spectra acquired from two different pixels (highlighted by crosses in the TEPL map) are shown in Fig. SI-6c. One of the MDH areas is highlighted by a rectangular box for which magnified CPD images were measured and shown in Fig. 4b and SI-6d. Discussion on CPD image and line profiles can be found in main text and in Fig. 4b,c. Gaussian fit to the line profile 1 gives a full width at half maxima of 13 ± 1 nm, which is a very good agreement to width of a single NPL (see sample 1 table S1). Using the dimensions of the NPL, CPD line profiles and TEPL map, we sketched the MDH area with five NPLs, as shown in Fig. SI-6d.

We also performed electrical characterization of this MDH using conductive AFM. The commercially available Cr/Au probes were used for the $I$–$V$ characterization. The schematic of the experimental configuration is shown in the inset of Fig. SI-6e. For the electrical characterization, we used bare Au substrate instead of Al$_2$O$_3$/Au used for the optical characterization. The representative $I$–$V$ curves of the sample acquired in four different areas are plotted in Fig. SI-6e. These four areas are marked in the AFM and CPD images of Fig. SI-6a, d. As expected, the $I$–$V$ curve for Area 4, i.e., on carbon residual (mainly PDMS), shows no current. The $I$–$V$ curve for Area 3 on WSe$_2$, on the other hand, exhibits a linear dependency with applied voltage. This ohmic-like behavior can be attributed to the highly efficient tunneling of the carriers through the ultra-thin ML-WSe$_2$ film. In contrast, $I$–$V$ curves for Areas 1 and 2 on MDH show rectification behaviour, a characteristic of type–II band alignment. The slightly higher conductivity and asymmetric $I$–$V$ characteristic of Area 2 was possibly because it was conducted on an isolated NPL interface, and the effective cross-section of the tip was larger than the MDH area; as such, it resulted in electric contact through the sides of the tip to the substrate.

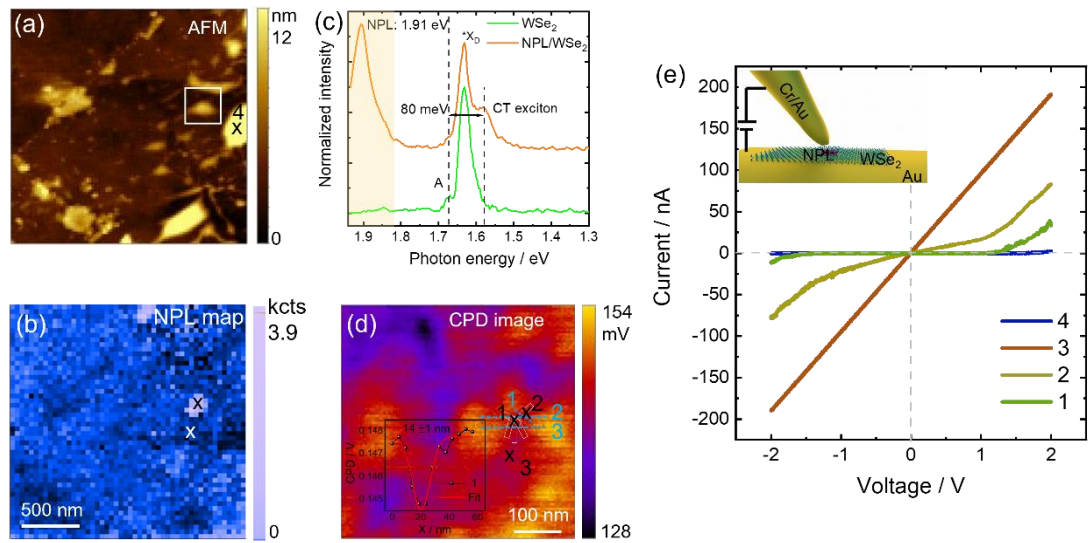

**Figure SI-6. Local electrical characterizations of MDH**. (a), (b) large-area AFM topography and corresponding TEPL map of NPL/WSe$_2$ MDH sample on bare Au. (c) Two representative TEPL spectra of the MDH system acquired from two spots as indicated in (b). TEPL map was created for the orange shaded area. (d) CPD image of the MDH area highlighted by a rectangle in (a). Three line profiles (blue dashed lines and numbers) were extracted and presented in Fig. 4c. Inset is the line profile 1 with gaussian fit. (e) *I–V* characterizations of the sample conducted at 4 different spots highlighted in (a) and (d) (black crosses and numbers). Inset is the schematic of the measurement setup.

## VII. Time-resolved PEEM of MDH

A simplified schematic of the time-resolved PEEM measurement of our sample is shown in Fig. SI-7a. For PEEM measurements, the sample discussed in Fig. 1 in the main text and SI-II was used. Two 20 fs pump pulses, one at 1.65 eV (near MoSe$_2$ A exciton) and the other at 1.51 eV (near the CT exciton energy), followed by a second probe pulse at 3.1 eV, were used for the measurements. The optical image of the sample is shown in Fig. SI-7b, and the respective PEEM microscope image is presented in Fig. SI-7c. Even though PEEM has a sub-micron spatial resolution, we collected and averaged the signal over the complete ML flake to increase sensitivity. The time-resolved PEEM spectra at the two pump excitations are discussed in Fig. 4h in the main manuscript.

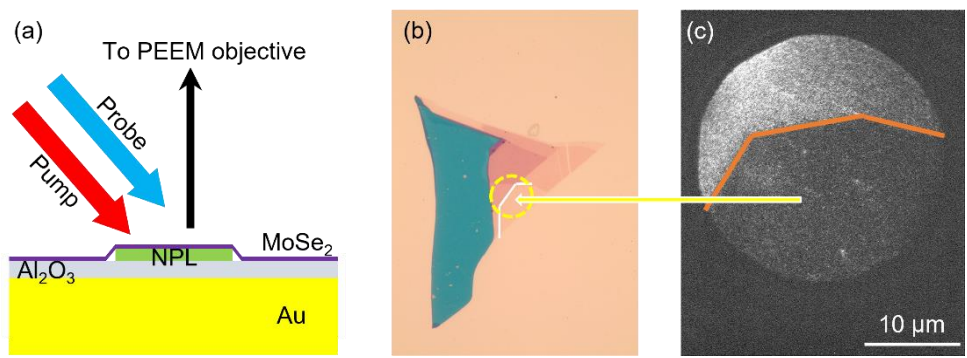

**Figure SI-7. Time-resolved PEEM of MDH**. (a) Simplified sketch of PEEM measurements on our MDH sample. (b) Optical image of the sample. (c) PEEM image of the region outlined by the circle in (b).

**Table S2: Fit parameters of ultrafast dynamics of tr-PEEM data**

$A_i$ and $\tau_i$ are the amplitude and time constants

| Fit parameters | At 1-65 eV | At 1.51 eV |
|---|---|---|
| A1 (%) | 56 | 49 |
| τ1 (fs) | 330 ± 30 | 140 ± 10 |
| A2 (%) | 44 | 51 |
| τ2 (fs) | 1100 ± (100) | 560 ± 30 |

## VIII. Time-resolved differential reflectance of MDH

We also measured exciton lifetime using time-resolved differential reflectance measurements. The output of a TiSapph (KM Labs, Griffen-10, $\lambda = 800$ nm, pump pulse < 20 fs) was split into a pump and probe path by a beam splitter and then focused onto the sample by a 40x Cassegrain objective. The pump fluence was ~ 4 mJ/cm$^2$. Fig. SI-8a shows the optical image of the sample. A differential reflectance map was acquired on the multilayer area of the sample (rectangular area in the optical image). The reflectance map is presented in Fig. SI-8b, which shows the NPL cluster at the centre. The time-resolved exciton decay curves on the cluster and the other outside of the cluster are shown in Fig. SI-8c. As can be seen, the off-cluster exciton lifetime is higher than the on-cluster exciton. Even though differential reflectance measurements were performed on multilayer MoSe$_2$, for which exciton dynamics are different than the monolayer MoSe$_2$, the overall exciton lifetime trend observed on- and off-cluster is consistent with our time-resolved PEEM results.

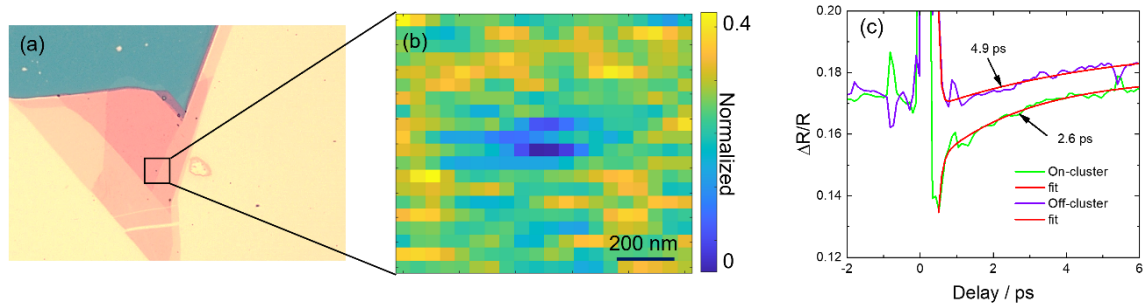

**Figure SI-8. Extraction of excitons lifetime**. (a) Optical image of the sample. (b) Differential reflectance mapping image of the AOI marked by the rectangle in (a). (c) Exciton decay curves measured on-cluster and off-cluster.

**References**

(1)  Carion, O.; Mahler, B.; Pons, T.; Dubertret, B. Synthesis, Encapsulation, Purification and Coupling of Single Quantum Dots in Phospholipid Micelles for Their Use in Cellular and in Vivo Imaging. *Nat. Protoc. 2007 210* **2007**, *2*, 2383–2390.

(2)  She, C.; Fedin, I.; Dolzhnikov, D. S.; Dahlberg, P. D.; Engel, G. S.; Schaller, R. D.; Talapin, D. V. Red, Yellow, Green, and Blue Amplified Spontaneous Emission and Lasing Using Colloidal CdSe Nanoplatelets. *ACS Nano* **2015**, *9*, 9475–9485.

(3)  Marino, E. Assembling Nanocrystal Superstructures, Universiteit van Amsterdam, 2019.

(4)  Marino, E.; Kodger, T. E.; Crisp, Y. W.; Timmerman, D.; Macarthur, K. E.; Arc Heggen, M.; Schall, P.; Marino, ] E; Kodger, T. E.; Timmerman, + D; Schall, P.; Risp, R. W. C.; Macarthur, K. E.; Heggen, M. Repairing Nanoparticle Surface Defects. *Angew. Chemie Int. Ed.* **2017**, *56*, 13795–13799.

(5)  Rossinelli, A. A.; Rojo, H.; Mule, A. S.; Aellen, M.; Cocina, A.; De Leo, E.; Schäublin, R.; Norris, D. J. Compositional Grading for Efficient and Narrowband Emission in CdSe-Based Core/Shell Nanoplatelets. *Chem. Mater.* **2019**, *31*, 9567–9578.

(6)  Hendricks, M. P.; Campos, M. P.; Cleveland, G. T.; Plante, I. J. La; Owen, J. S. A Tunable Library of Substituted Thiourea Precursors to Metal Sulfide Nanocrystals. *Science (80-. ).* **2015**, *348*, 1226–1230.

(7)  Darlington, T. P.; Carmesin, C.; Florian, M.; Yanev, E.; Ajayi, O.; Ardelean, J.; Rhodes, D. A.; Ghiotto, A.; Krayev, A.; Watanabe, K.; Taniguchi, T.; Kysar, J. W.; Pasupathy, A. N.; Hone, J. C.; Jahnke, F.; Borys, N. J.; Schuck, P. J. Imaging Strain-Localized Excitons in Nanoscale Bubbles of Monolayer WSe2 at Room Temperature. *Nat. Nanotechnol. 2020 1510* **2020**, *15*, 854–860.

(8) Rahaman, M.; Selyshchev, O.; Pan, Y.; Schwartz, R.; Milekhin, I.; Sharma, A.; Salvan, G.; Gemming, S.; Korn, T.; Zahn, D. R. T. Observation of Room-Temperature Dark Exciton Emission in Nanopatch-Decorated Monolayer WSe2 on Metal Substrate. *Adv. Opt. Mater.* **2021**, 2101801.

(9) Empedocles, S. A.; Bawendi, M. G. Quantum-Confined Stark Effect in Single CdSe Nanocrystallite Quantum Dots. *Science (80-. ).* **1997**, *278*, 2114–2117.